\documentclass[11pt]{article}
\usepackage{moriond,epsfig,graphicx}

\def\mgg{M_{\gamma\gamma}}
\def\lsim{\buildrel < \over {_\sim}}
     
\def\pt{p_{\rm T}}
\def\et{E_{\rm T}}
\def\etmax{E_{\rm T\,max}}

\def\be{\begin{equation}}
\def\ee{\end{equation}}
\def\bea{\begin{eqnarray}}
\def\eea{\end{eqnarray}}

\def\Rjet{R_{\rm jet}}
\begin{document}
\vspace*{-0.3cm}
{\small MSUHEP--030611 \hfill UCLA/03/TEP/17 \hfill SLAC--PUB--9954} \\
\vspace*{-0.3cm}

\title{Di-photon production at the LHC}

\author{ Zvi Bern${}^{a}$,
 Lance Dixon${}^{b}$,
 and Carl Schmidt${}^{c}$ (presenter)
 }

\address{${}^{a}$Department of Physics and Astronomy,
                   UCLA, Los Angeles, CA 90095-1547, USA\\
${}^{b}$Stanford Linear Accelerator Center,
                          Stanford University, Stanford, CA 94309, USA\\
${}^{c}$Department of Physics and Astronomy,
Michigan State University, East Lansing, MI 48824, USA}

\maketitle\abstracts{
The standard model production of two photons is one of the most
important backgrounds to light Higgs boson production at the LHC.  In this
talk we discuss the di-photon background, with emphasis on the
effects of the recently calculated next-to-leading-order (NLO) corrections
to the gluon-gluon-initiated component.  We find that the $K$-factor for this
component is
smaller than that for the analogous $gg\rightarrow H$ process, and that the correction
to the total irreducible di-photon production is modest.  We also investigate
ways to enhance the statistical significance of the Higgs signal in the
$\gamma\gamma$ channel.}

\section{Introduction}
One of the primary goals of the LHC is to provide clues to the nature
of electroweak symmetry breaking.  In the standard model this symmetry 
breaking is provided by the condensation of a fundamental weak scalar 
doublet, which leaves behind a single neutral scalar, the Higgs boson, as
residue.  This standard model Higgs boson is constrained
by precision electroweak measurements to have $m_H \lsim
196$--$230$ GeV at 95\% CL.\cite{HiggsRadCorr}  A light neutral scalar boson
with similar properties is also often predicted in extensions of the
standard model; for example, in the Minimal Supersymmetric Standard Model 
(MSSM) the lightest Higgs boson is predicted to have a mass below about 
135 GeV.~\cite{SusyHiggs}  

For $m_H < 140$ GeV, the most promising discovery
mode for the Higgs boson at the LHC 
involves production via gluon fusion, $gg\to H$, 
followed by the rare decay into two photons,
$H\to\gamma\gamma$.~\cite{HggVertex}  Although this
mode has a very large continuum $\gamma\gamma$
background,\cite{HBkgdgammagamma} the excellent mass resolution of
the LHC detectors should allow the detection of the narrow Higgs
resonance signal above the background.\cite{ATLAS}  
For optimizing this analysis, 
it is necessary to have the best theoretical 
understanding of both signal and background beforehand.  

The perturbative contribution to $pp \to \gamma\gamma X$ proceeds at lowest 
order via the quark annihilation subprocess $q\bar q \to \gamma\gamma$.
The NLO corrections to this process have been 
calculated,\cite{TwoPhotonBkgd1,DIPHOX} as have processes involving
parton fragmentation to photons at NLO.\cite{DIPHOX}  
The remaining
important perturbative contribution is the gluon annihilation subprocess 
$gg \to \gamma\gamma$.  Although this arises at order $\alpha_s^2$
through a quark box loop, it is still of the same size as the
quark annihilation subprocess, due to the large gluon luminosity at the 
LHC.\cite{HBkgdgammagamma,ADW,TwoPhotonBkgd1,DIPHOX}
All of these perturbative contributions have been included in the Monte
Carlo program {\tt DIPHOX},\cite{DIPHOX} with the direct and fragmentation
contributions evaluated at NLO and the gluon box diagram evaluated at
lowest order $\alpha_s^2$.  In this talk we report on the impact 
of extending the gluon box contribution to NLO.~\cite{bds}  (Note that we
refer to this correction as NLO in comparison to the lowest-order
gluon box contribution; in fact, it is N$^3$LO compared to the
order $\alpha_s^0$ quark annihilation process.)

Before we discuss this correction and some phenomenological studies
in the next sections,
it is important to emphasize that our analysis only includes the
irreducible di-photon background.  There is also a very large
reducible background, arising from photons which are faked by jets
or hadrons, especially $\pi^0$'s.  Although these reducible backgrounds 
can be suppressed greatly by photon isolation 
cuts,\cite{ATLAS} more detailed
experimental studies of pion fragmentation at large momentum fraction
in jets are necessary to accurately quantify the impact of this 
reducible background.\cite{PiBkgd}

\section{Effect of gluon fusion at NLO}

Although the gluon box contribution 
begins at one-loop and is order $\alpha_s^2$, the QCD corrections
to it
can be treated exactly as a NLO calculation,\cite{GGGamGam}
independent from the other perturbative contributions at this order.
We have implemented this NLO cross section in a Monte Carlo routine,
which allows the imposition of general kinematic cuts.  Details
of this calculation can be found in our full journal article.\cite{bds}


For the phenomenological analyses, 
we impose the following cuts on the two photons:
$\pt(\gamma_1) > 40\hbox{ GeV}$, $\pt(\gamma_2) > 25\hbox{ GeV}$,
$| y(\gamma_{1,2}) | < 2.5$.
In addition, we impose one of two photon isolation criteria:
{\it standard} cone isolation --- the amount of transverse hadronic
energy $\et$ in a cone of radius 
$R = \sqrt{(\Delta\eta)^2 + (\Delta\phi)^2}$ must be less than $\etmax$;
or 
{\it smooth} cone isolation~\cite{Frixione} 
--- the amount of transverse hadronic energy $\et$ in {\it all} cones 
of radius $r$ with $r < R$ must be less than 
$\etmax(r) \equiv 
\pt(\gamma) \, \epsilon \, ( 1 - \cos r)/( 1 - \cos R )$,
%
for some $\epsilon$.
The smooth cone isolation criterion is designed to remove all
fragmentation contributions in an infrared-safe manner.  For all calculations
using this isolation criterion we implemented the NLO quark-fusion 
contributions, as well as the NLO gluon box contributions, 
directly in our Monte Carlo.  For calculations using the standard cone 
isolation criterion we have used our own Monte Carlo to calculate the
NLO gluon box contributions, and we have used {\tt DIPHOX}~\cite{DIPHOX}
 to calculate all other contributions.

\begin{figure*}[t]
\includegraphics[width=7.0cm]{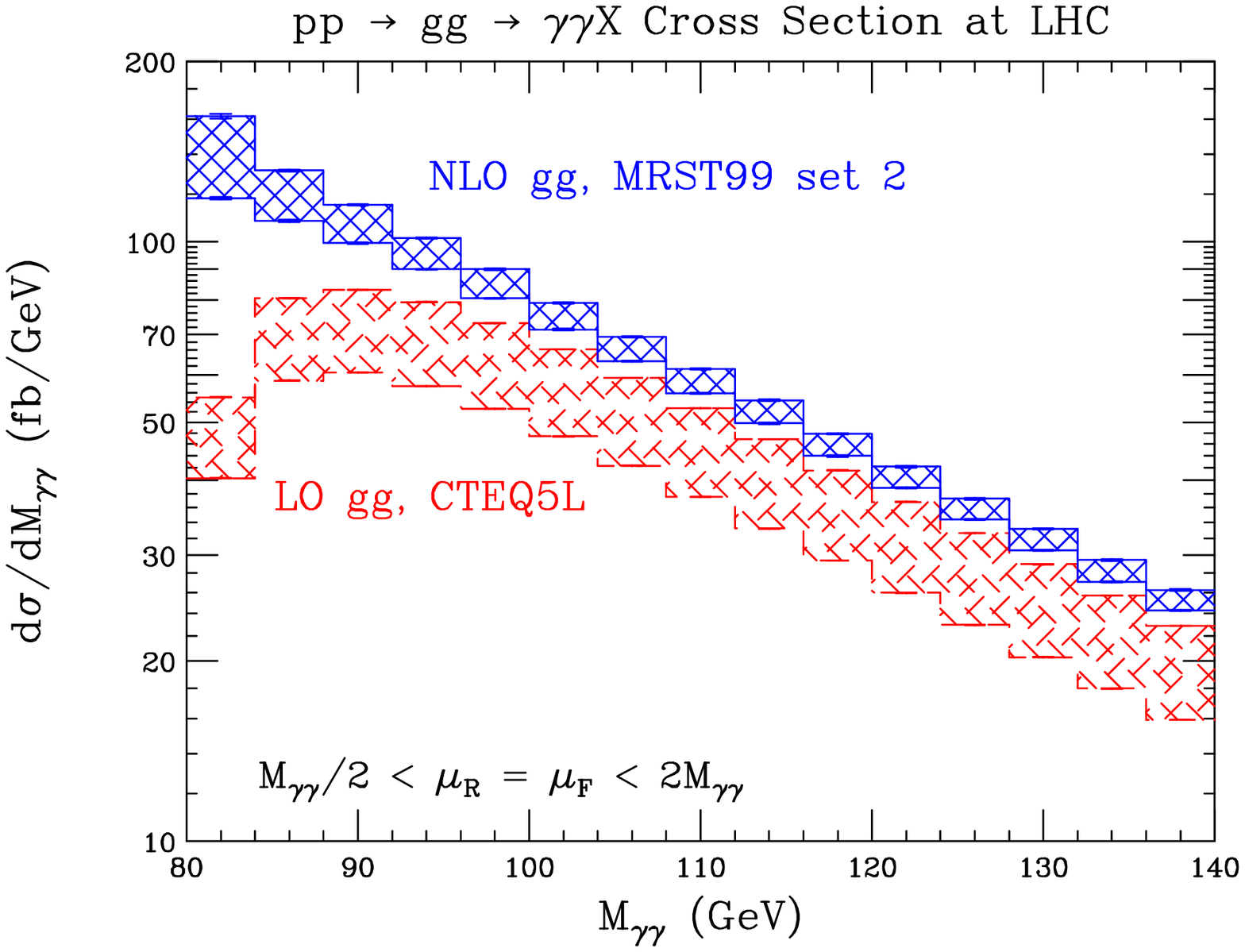}
\includegraphics[width=7.0cm]{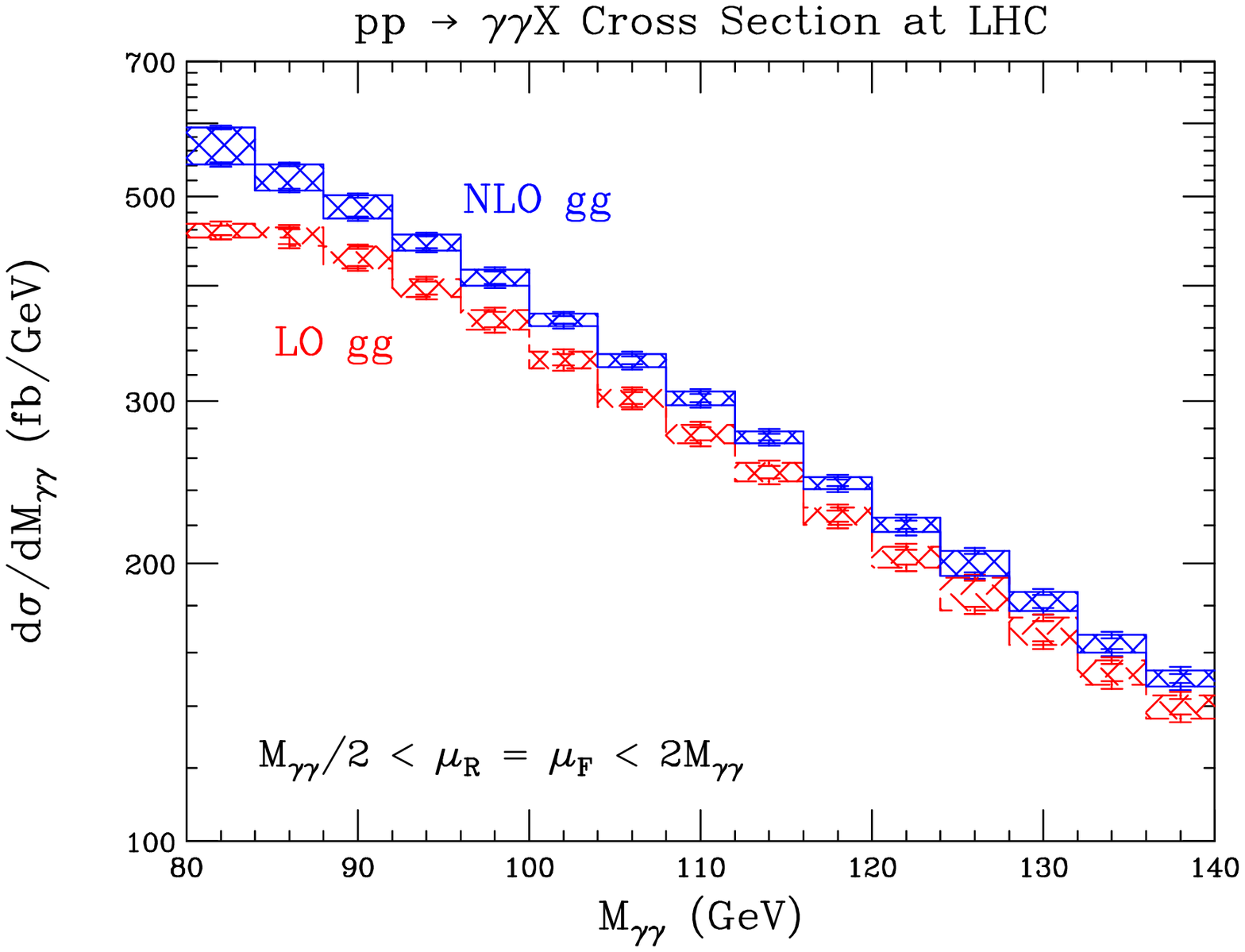}
\caption{\label{fig34_rp4_diag} Scale dependence of (a) the gluon fusion
subprocess contribution to $pp \to \gamma\gamma X$, and (b) the total 
$pp\to \gamma\gamma X$ production cross section, for standard photon
isolation with $R=0.4$, $\etmax = 15$~GeV.  In both plots, the
bands represent the result of varying $\mu_R$ and $\mu_F$ together
over region $0.5 \mgg < \mu_R = \mu_F < 2 \mgg$.  The dashed (solid) hatched
band corresponds to including the gluon fusion subprocess at LO (NLO).
For the leading order band in (a) only, the LO CTEQ5L
 parton distributions$^{14}$
were used; otherwise the NLO MRST99 
set 2 distributions$^{13}$
were employed.
\hfill\  }
\end{figure*}

In fig.~\ref{fig34_rp4_diag}(a) we plot just the gluon box contribution
at LO and NLO while varying the renormalization and factorization
scales together over the range $0.5 \mgg < \mu_R = \mu_F < 2 \mgg$.
As is typical of gluon-initiated processes, the NLO cross section 
is larger than at LO, and the scale-dependence is reduced.  We note,
however, that the reduction in scale-dependence
is much less impressive if $\mu_R$
and $\mu_F$ are allowed to vary independently.\cite{bds}  

It is useful to compare the NLO enhancement in the $gg\to\gamma\gamma X$
process with that of the $gg\to HX$ process by comparing $K$ factors, defined
as the ratio of the NLO and LO cross sections.  Using the NLO MRST99 
set 2 distributions~\cite{MRST99} 
for both the numerator and denominator with $\mu_R = \mu_F
=0.5 \mgg$ and the same sets of cuts as in fig.~\ref{fig34_rp4_diag}(a)
for $\mgg=118$ GeV, we obtain $K_{gg\to\gamma\gamma}=1.61$
and $K_{\rm Higgs}=2.54$.  Thus, some earlier studies of the 
di-photon background, which adopted the $K$ factor for Higgs production,
overestimated this background.  The larger $K$ factor for Higgs production
can be traced to two effects. The effective $Hgg$ coupling, which
results from the heavy top quark loop, receives a short-distance 
renormalization that has no counterpart in direct $\gamma\gamma$
production.  In addition, the real correction to Higgs production
has a harder transverse momentum spectrum than direct $\gamma\gamma$
production, since the dominant scale in the loop is the top mass, rather
than the parton momenta.  
Recently, the higher order 
coefficients that affect the $K$ factor in the resummation of the
di-photon cross section at small transverse momentum have been 
extracted from this NLO calculation.\cite{pavel}

In fig.~\ref{fig34_rp4_diag}(b) we show the effects of computing the
gluon fusion subprocess at NLO on the total irreducible di-photon
background.  The upper band contains all processes, including
gluon fusion, quark annihilation, and fragmentation contributions at NLO,
while the lower band contains gluon fusion at LO and the other processes
at NLO.  The scale variation and kinematic cuts are the same as
for fig.~\ref{fig34_rp4_diag}(a).  As before, the scale dependence is
much greater for both bands if $\mu_R$ and $\mu_F$ are varied
independently.\cite{bds}  For $\mgg$ greater than 100 GeV,
the NLO gluon box correction to the total cross section is about
10\% or less; therefore this subprocess can be considered to be
under adequate theoretical control.

\section{Di-photon background kinematics and Higgs Signal}

Using our Monte Carlo, we have studied some kinematic properties of the 
di-photon production 
with the aim of increasing the statistical significance of the Higgs
signal above background.  For this study we have implemented
the Higgs signal in the Monte Carlo at NLO in the heavy top mass limit, 
with subsequent decay to $\gamma\gamma$.  We assumed 
a Higgs mass of 118 GeV, and we counted the number of events in a mass 
bin of 4 GeV for $30$ fb$^{-1}$ of integrated luminosity.  We also
included an experimental efficiency factor of 0.57 for both signal
and background (0.81 per $\gamma$ for identification, 0.87 for fiducial
cuts), and we included, as a rough estimate, a reducible background of 
20\% of the $\gamma\gamma$ continuum background.  

One feature that distinguishes the Higgs signal from the irreducible
background is the correlation between the photons and any 
accompanying hadronic radiation.  Since the Higgs boson is a colorless
object, its decay photons will be produced uncorrelated with hadronic
radiation.  On the other hand, a significant component of the background
consists of $qg\to qg\gamma$, where the $\gamma$ has a collinear 
enhancement when produced near the final-state quark.  This suggests
that one can suppress the background relative to the signal by
increasing the severity of the isolation cut---either by increasing
the cone size $R$, or decreasing $\epsilon$ or $\etmax$.  As an example
the statistical significance ($S/\sqrt{B}$) was increased by 7\%
by increasing $R$ from 0.4 to 2, while keeping $\etmax=15$ GeV fixed.
Similar results were found for the smooth cone algorithm.

Unfortunately, an isolation cone as large as $R=2$ may not be 
phenomenologically viable for both theoretical and experimental reasons.  
A more infrared-safe and experimentally better-behaved procedure is to 
veto on jets within a larger cone $\Rjet$ around the photons, in addition 
to an isolation cut with $R<\Rjet$.  The jet veto is more infrared-safe
than a large isolation cone, because it only restricts hadronic energy 
within the jet cone, not the full jet veto region $\Rjet$.  It is also
more viable experimentally, since one need not worry about loss of
efficiency due to detector noise or nonperturbative sources of hadronic 
energy in the large isolation cone.  We have investigated the use of
a jet veto to enhance the statistical significance of the Higgs signal,
but found the significance to be relatively insensitive to it.  More
study is needed to find the best way to utilize the hadronic energy
distribution in the events to optimize the Higgs signal.

A second feature that distinguishes the signal and background is the
angular distribution of the photons.  The signal photons, coming from the
decay of a scalar particle, are isotropic in the Higgs rest frame, whereas
the background photons tend to be more peaked along the beam axes.  
This results in a distinctly different distribution in the 
di-photon rapidity difference,
$y^* = (y(\gamma_1) - y(\gamma_2))/2$, for the signal and background events.  
This distribution is the
most robust discriminator that we found, offering a modest ($\sim4$\%) 
improvement in the statistical significance of the Higgs signal.

More details of this calculation and the analysis 
can be obtained in our full journal article.\cite{bds}
Further studies, which include reducible di-photon background 
contributions,\cite{PiBkgd} additional  NNLO contributions, and more detailed
experimental simulations will greatly enhance our understanding
of the di-photon background, leading to 
 increased sensitivity for
the Higgs search at the LHC, and potentially reducing the time 
required for the discovery of a light Higgs boson.


\section*{Acknowledgments}
We thank Thomas Binoth for providing us with a copy of {\tt DIPHOX}.
The work of Z.B. and L.D. was supported by the US Department of Energy 
under contracts DE-FG03-91ER40662 and DE-AC03-76SF00515, respectively.  
The work of C.S. was supported by the US National Science Foundation under
grant PHY-0070443.


\section*{References}
\begin{thebibliography}{99}
\bibitem{HiggsRadCorr}
G.~Degrassi,
arXiv:hep-ph/0102137; 
J.~Erler,
arXiv:hep-ph/0102143;
D. Abbaneo {\it et al.} [ALEPH, DELPHI, L3 and OPAL Collaborations, 
LEP Electroweak Working Group, and SLD Heavy Flavor and Electroweak Groups],
arXiv:hep-ex/0112021.


\bibitem{SusyHiggs}
M.~Carena {\it et al.},
Nucl.\ Phys.\ B {\bf 580}, 29 (2000);
J.R.~Espinosa and R.~Zhang,
Nucl.\ Phys.\ B {\bf 586}, 3 (2000);
A.~Brignole {\it et al.},
Nucl.\ Phys.\ B {\bf 631}, 195 (2002).

\bibitem{HggVertex}
J.R.~Ellis {\it et al.}, 
Nucl.\ Phys.\ B {\bf 106}, 292 (1976);
M.A.~Shifman {\it et al.}, 
Sov.\ J.\ Nucl.\ Phys.\  {\bf 30}, 711 (1979)
[Yad.\ Fiz.\  {\bf 30}, 1368 (1979)];
J.F.~Gunion {\it et al.},
Phys.\ Rev.\ D {\bf 34}, 101 (1986); 
J.F.~Gunion, G.L.~Kane and J.~Wudka,
Nucl.\ Phys.\ B {\bf 299}, 231 (1988).

\bibitem{HBkgdgammagamma}
R.K.~Ellis, I.~Hinchliffe, M.~Soldate and J.J.~van der Bij,
Nucl.\ Phys.\ B {\bf 297}, 221 (1988).

\bibitem{ATLAS}
ATLAS collaboration,
``ATLAS detector and physics performance, technical design report,'' 
vol. 2, report CERN/LHCC 99-15, ATLAS-TDR-15;
CMS collaboration,
``CMS: The electromagnetic calorimeter, technical design report,''
report CERN/LHCC 97-33, CMS-TDR-4;
V.~Tisserand, 
``The Higgs to two photon decay in the ATLAS detector,''
talk given at the VI International Conference on Calorimetry in
High-Energy Physics, Frascati (Italy), June, 1996, LAL 96-92; Ph.D. thesis,
LAL 97-01, February, 1997;
M.~Wielers, ``Isolation of photons,'' report ATL-PHYS-2002-004.

\bibitem{TwoPhotonBkgd1}
E.L.~Berger {\it et al.}, 
Nucl.\ Phys.\ B {\bf 239}, 52 (1984); 
P.~Aurenche {\it et al.},
Z.\ Phys.\ C {\bf 29}, 459 (1985); 
B.~Bailey {\it et al.},
Phys.\ Rev.\ D {\bf 46}, 2018 (1992); 
B.~Bailey and J.F.~Owens,
Phys.\ Rev.\ D {\bf 47}, 2735 (1993); 
B.~Bailey and D.~Graudenz,
Phys.\ Rev.\ D {\bf 49}, 1486 (1994);
C.~Balazs {\it et al.},
Phys.\ Rev.\ D {\bf 57}, 6934 (1998);
C.~Balazs and C.-P.~Yuan,
Phys.\ Rev.\ D {\bf 59}, 114007 (1999)
[Erratum-ibid.\ D {\bf 63}, 059902 (1999)];
T.~Binoth {\it et al.},
Phys.\ Rev.\ D {\bf 63}, 114016 (2001);
T.~Binoth,
arXiv:hep-ph/0005194.

\bibitem{DIPHOX}
T.~Binoth {\it et al.},
Eur.\ Phys.\ J.\ C {\bf 16}, 311 (2000).

\bibitem{ADW}
L.~Ametller, E.~Gava, N.~Paver and D.~Treleani,
Phys.\ Rev.\ D {\bf 32}, 1699 (1985); 
%
D.A.~Dicus and S.S.D.~Willenbrock,
Phys.\ Rev.\ D {\bf 37}, 1801 (1988).

\bibitem{bds}
Z.~Bern, L.~Dixon and C.~Schmidt,
Phys.\ Rev.\ D {\bf 66}, 074018 (2002).

%
\bibitem{PiBkgd}
T.~Binoth, J.P.~Guillet, E.~Pilon and M.~Werlen,
arXiv:hep-ph/0203064.

\bibitem{GGGamGam}
Z.~Bern, A.~De~Freitas and L.J.~Dixon,
JHEP {\bf 0109}, 037 (2001);
Z.~Bern, L.~Dixon and D.A.~Kosower,
Phys.\ Rev.\ Lett.\  {\bf 70}, 2677 (1993);
D.~de~Florian and Z.~Kunszt,
Phys.\ Lett.\ B {\bf 460}, 184 (1999);
%
C.~Balazs {\it et al.},
Phys.\ Lett.\ B {\bf 489}, 157 (2000).


\bibitem{Frixione}
S.~Frixione,
Phys.\ Lett.\ B {\bf 429}, 369 (1998).

\bibitem{MRST99}
A.D.~Martin {\it et al.},
Eur.\ Phys.\ J.\ C {\bf 14}, 133 (2000).

\bibitem{CTEQ}
H.L.~Lai {\it et al.}  [CTEQ Collaboration],
Eur.\ Phys.\ J.\ C {\bf 12}, 375 (2000).


\bibitem{pavel}
P.~M.~Nadolsky and C.~R.~Schmidt,
Phys.\ Lett.\ B {\bf 558}, 63 (2003).
\end{thebibliography}
\end{document}